# Chemical vapor deposited graphene – From synthesis to applications

S. Kataria[1], S. Wagner[1], J. Ruhkopf[1], A. Gahoi[1], H. Pandey[1], R. Bornemann[2], S. Vaziri[3], A.D. Smith[3], M. Ostling[3], M.C. Lemme[1,*]

[1]University of Siegen, Graphene-based Nanotechnology, Hölderlinstr. 3, 57076 Siegen, Germany

[2]University of Siegen, High Frequency and Quantum Electronics, Hölderlinstr. 3, 57076 Siegen, Germany

[3]KTH Royal Institute of Technology, School of ICT, Electrum 229, 16440 Kista, Sweden

**Abstract**

Graphene is a material with enormous potential for numerous applications. Therefore, significant efforts are dedicated to large-scale graphene production using a chemical vapor deposition (CVD) technique. In addition, research is directed at developing methods to incorporate graphene in established production technologies and process flows. In this article, we present a brief review of available CVD methods for graphene synthesis. We also discuss scalable methods to transfer graphene onto desired substrates. Finally, we discuss potential applications that would benefit from a fully scaled, semiconductor technology compatible production process.

**Keywords**: Graphene, chemical vapor deposition, graphene transfer, hot electron transistor, pressure sensor

*Corresponding author:
max.lemme@uni-siegen.de
Phone: +49 (0) 271 740 4035
Fax: +49 (0) 271 10435

# 1. Introduction

Graphene, the "first" of many two-dimensional materials, has received tremendous attention over the years since its discovery in 2004 [1]. Graphene is endowed with many extraordinary intrinsic properties that include high carrier mobility [2], broadband optical absorption [3], high current density [4], tensile strength in excess of 1 TPa [5] and high thermal conductivity [6]. Based on these properties, graphene is considered for numerous applications in various technological areas [7] covering microelectronics [8], optoelectronics [9] and nanoelectromechanical systems (NEMS) [10]. In addition, the fact that graphene is largely compatible with current complementary metal oxide semiconductor (CMOS) technology means that the material can be easily adopted for graphene co-integration with silicon devices [11]. Particular promise is seen for integration at the back-end-of-the-line, where graphene could provide added functionality to the existing CMOS platform [12]. With all this in mind, most or all of the potential applications require large-scale production of graphene layers with uniform thickness.

The highest quality graphene samples can be obtained by mechanical exfoliation of graphite [1]. However, this method yields only micron sized graphene flakes. While these are excellent for fundamental research, their thickness, size and uniformity is not controllable. Epitaxial graphene can be grown on a large scale by thermal decomposition of silicon carbide (SiC) [13]. The graphene grown by this method exhibits good electronic properties and is suitable for large-scale processing of graphene field effect devices grown directly on SiC substrates. It is, however, difficult to transfer the graphene to other substrates. In addition, the high process temperatures, the high cost of SiC substrates and their limited size scalability restrict its co-integration with silicon CMOS technology. Chemical vapor deposition (CVD) is a potential method for large-scale production of graphene. A comparison of different production methods, mentioned here, is shown in Table 1. After the first reports of CVD growth of graphene, the method has emerged as a potential pathway for graphene commercialization [14,15]. The most commonly used growth substrate is copper (Cu), because graphene growth on Cu can be limited to monolayer thickness with excellent uniformity over a large area.

In this paper, we present an overview of different variants of CVD methods for graphene growth on Cu substrates. Most of the graphene applications require its deposition on dielectric substrates. For this, CVD graphene on Cu needs to be transferred to respective substrates. Therefore, we also present an overview of existing graphene transfer methods with their advantages and disadvantages. Finally, some potential applications of CVD graphene are also discussed.

**Table 1** Comparison of graphene production methods

| Method | Crystallite size | Carrier mobility on $SiO_2$ /SiC at RT ($cm^2$/V.s)* | Applications | Remarks |
|---|---|---|---|---|
| **Mechanical exfoliation** | 100 μm | > 10,000 [1] | Fundamental research | No potential for industry scale |
| **SiC decomposition** | 50 μm | > 10,000 [16] | High frequency electronic devices and monolithic integration | Scalable to SiC substrate size, Non-transferrable |
| **CVD** | ~ 5 mm [17] | ~ 16,000 [17] | Biosensing, nanoelectronics, photonics, transparent conducting layers | Ultimately scalable, Transferrable to desirable substrates |

*The mobility values are for supported graphene at ambient conditions.

## 2. CVD growth of graphene

### 2.1 Thermal CVD

In a typical CVD process, appropriate precursor species are fed into a reaction chamber. In the chamber, chemical reactions take place leading to the production of solid material on a substrate kept at elevated temperatures. In case of graphene growth, a hydrocarbon gas such as $CH_4$ is used as a precursor to yield graphitic material on a suitable catalytic material. The main steps of CVD graphene growth are the decomposition of hydrocarbon into carbon and then the formation of the graphitic structure i.e. graphene on the catalytic surface. Therefore, the most important roles of a catalyst in graphene growth are to lower the energy barrier of reactions involving hydrocarbon pyrolysis and effective formation of graphene layers. This condition is mostly satisfied by transition metals like Ni, Cu, Pt, Pd, Rh, Fe and Co. Their catalytic activity is argued to arise from partially filled d orbitals [18]. In the case of Ni, once the hydrocarbon is decomposed into carbon atoms, the carbon can diffuse into the bulk of the transition metal and later precipitate to the surface during cooling. Similar processes take place for other carbide forming metals like Fe and Co. If the carbon solubility in the metal is limited or negligible, then graphene formation takes place as a surface process, e.g. in the case of Cu and Ru. It has been demonstrated experimentally that graphene growth on Ni is diffusion-precipitation dominated while it is a surface diffusion process on Cu surfaces [15]. Therefore, it is difficult to control uniformity and thickness of graphene layers on Ni and similar metals. In contrast, graphene growth is restricted mostly to monolayer thickness on Cu and similar surfaces, because once a monolayer is formed, the graphene film acts as a diffusion barrier for other carbon atoms. Therefore, and because of the wide availability and low cost, Cu is the mostly used catalyst for CVD graphene growth.

There are many ways through which hydrocarbons can be decomposed or pyrolysed into carbon atoms. One of the most commonly used methods is by high temperatures and the process is therefore termed as thermal CVD. For large area thermal CVD growth of graphene, $CH_4$ gas is decomposed at elevated temperatures of around 1000°C. Fig. 1(a) shows the schematic of a typical thermal CVD hot wall reactor. In our experiments, a rapid thermal processing cold wall reactor (Moorfield nanoCVD) has been used, where only the Cu foil is heated through a heating stage (Fig. 1(b)). Cu foil is heated to high temperatures in an atmosphere of Ar and $H_2$ gases. The substrate is first annealed for a certain length of time to reduce the oxides on the Cu surface and for grain growth. For graphene growth, an appropriate mixture of $CH_4/H_2$ gas is then used. During this time, $CH_4$ dissociates into radicals/atoms and surface reactions involving adsorption and surface diffusion take place at the Cu surface leading to graphene growth, as shown schematically in Fig. 1(c).

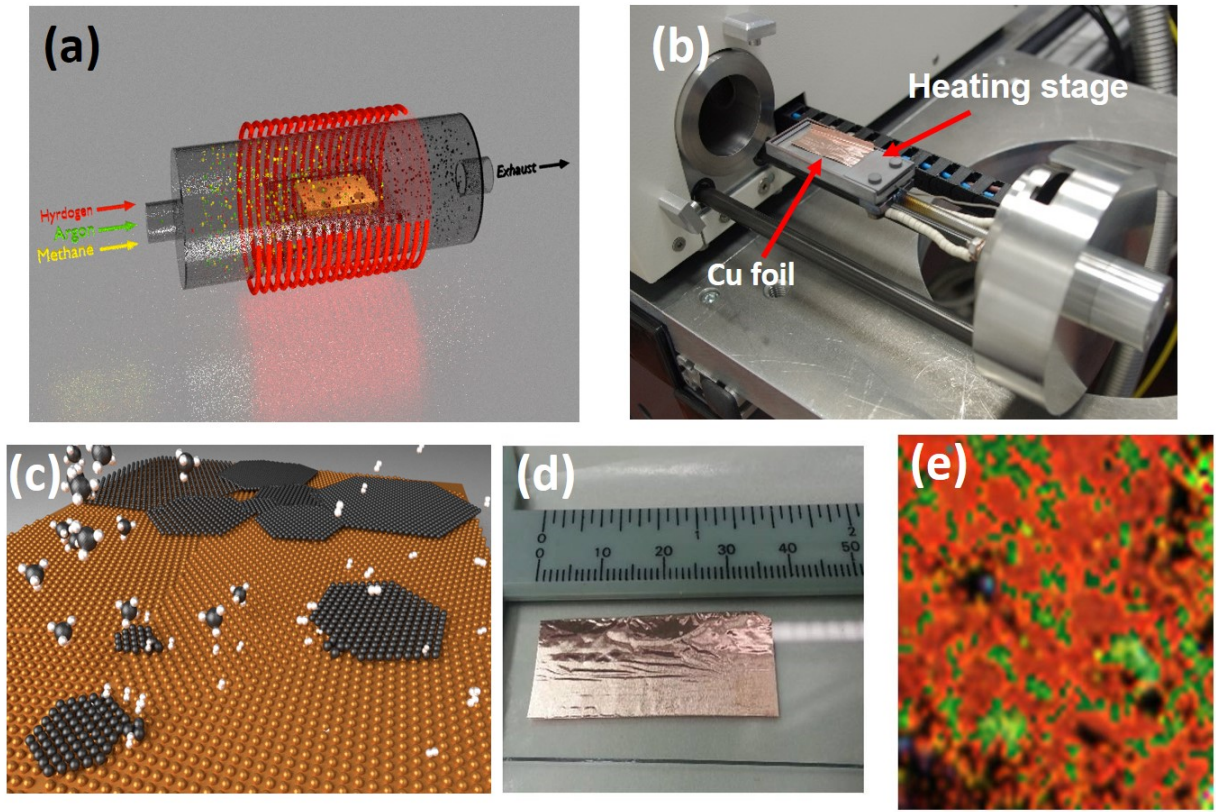

Figure 1. (a) Schematic of typical CVD of graphene using $CH_4$ gas as a precursor. (b) Photograph of a Cu foil kept on a heating stage before loading in a rapid thermal processing cold wall reactor (c) Graphene growth on Cu as a surface reaction. (d) CVD

graphene on Cu foil. (e) A Raman map of graphene film transferred on $SiO_2$/Si substrate. The area is $64 \times 64$ μm$^2$. Red color in the image represents single layer coverage. Green color represents multilayer regions.

Hydrogen helps to reduce the native surface oxides on the Cu surface, which passivates the growth of graphene. Recently, the role of surface oxygen on graphene growth has been investigated thoroughly and it has been found that surface oxygen is helpful in growing large grained graphene on Cu surface [19]. This happens because surface oxides reduce the nucleation density of graphene nuclei, thereby encouraging the large growth of single grains. The thermal CVD process can be carried out at pressures ranging from low pressures to atmospheric pressure. Fig. 1(d) shows the photograph of a Cu foil after graphene growth. The quality of graphene is assessed by Raman area mapping after transfer to a $SiO_2$/Si substrate, as shown in Fig. 1(e).

The main parameters that affect the growth of graphene on Cu are process pressure and $CH_4$ to $H_2$ ratio. The role of hydrogen during thermal CVD has been demonstrated and it was found that $H_2$ partial pressure has a significant effect on the graphene grain size and shape [20]. Generally, graphene growth is limited to monolayer on Cu but by changing the working pressure and $H_2/CH_4$ ratio, researchers are now able to control the number of graphene layers on Cu by thermal CVD [21]. The proposed growth mechanism involve formation of graphene seeds, which pre-define the thickness of graphene layers, by supersaturated surface carbon in Cu. These graphene seeds grow independently before forming continuous graphene film. The growth mechanism is proposed to be simultaneous-seeding and self-limiting process rather than a layer-by-layer growth. Direct growth of graphene on dielectric substrates has been demonstrated using Cu thin films as catalysts. Recently, graphene has been grown using Cu vapors by placing a $SiO_2$ substrate directly underneath a Cu foil [22]. Roll to roll production of graphene films has been demonstrated using thermal CVD making it a promising technique for continuous large-area growth of graphene [23]. The processing time of thermal CVD is around 2 to 3 hours including heating, annealing, growth and cooling times. This can vary depending on the annealing time and $CH_4$ exposure time for graphene growth. In conventional thermal CVD, the substrate is heated to the desired temperature by convection or radiation heating and takes a longer time to reach the deposition temperature. However, the heating and cooling time can be reduced dramatically by using rapid thermal annealing (RTA) systems. This reduces the total processing time to about 30 to 40 minutes.

Generally, the graphene grown on Cu foils is polycrystalline in nature and contains grain boundaries that degrade their electronic properties. To circumvent the problem of polycrystallinity, efforts have been made to grow large-area single crystal graphene. For this, single crystal germanium (Ge) has been used as a catalytic substrate, utilizing the low solubility of C in Ge even at its melting point [24–26]. Wang *et al.* have demonstrated thermal CVD growth of uniform and wafer scale graphene on Ge substrate [24]. By depositing graphene at different temperatures, they found that graphene growth on Ge is self-limited and surface mediated as in case of Cu. However, graphene transfer from Ge substrate to desired substrates led to wrinkled and folded regions. Lippert *et al.* have also reported direct uniform deposition of graphene on Ge (001) layers on Si (001) wafers [25]. Their results indicate that graphene can be directly deposited at the active regions of Si transistors with a process compatible with Si technology. Lee *et al.* have reported the growth of wafer scale growth of single crystal monolayer graphene on hydrogen terminated Ge substrates [26]. Here, a controlled low pressure thermal CVD process leads to wrinkle free growth on epitaxially grown single crystal Ge layers on Si (110) wafers. Weak adhesion between the graphene and an H terminated Ge buffer layers allowed mechanical exfoliation of graphene using Au thin films as a support layer. Using this transfer process, the authors demonstrated that Ge substrates can be reused several times for graphene growth. In addition to the potential quality and cost advantage, the growth of graphene on Ge substrates may solve contamination issues associated with the use of catalytic metals in conventional graphene CVD.

**2.2 Alternative CVD methods**

A hot filament CVD (HF-CVD) technique has been used to grow diamond films and carbon nanotubes on industrial scale. In this technique, the hydrocarbon gas like $CH_4$ is decomposed by thermal means using a hot filament, as shown in Fig. 2(a). Recently, the technique has been used to grow high quality graphene on Cu foils [27,28]. Fig. 2(b) – 2(e) show the quality of graphene grown by HF-CVD technique analyzed by optical and Raman spectroscopy.

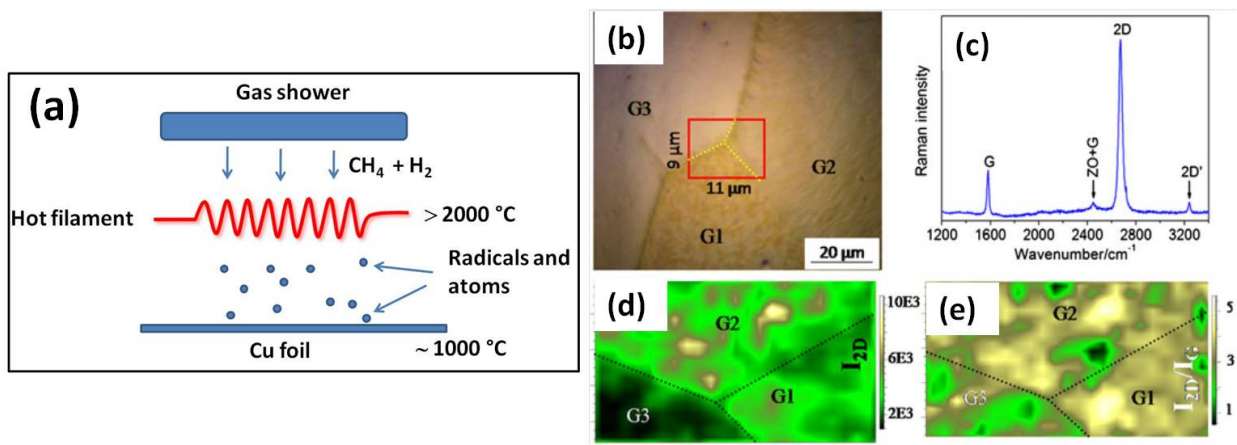

Figure 2. (a) Schematic depicting the CVD process in a Hot-Filament CVD set up. (b) Optical micrograph of a Cu foil after graphene growth depicting the three grains G1, G2 and G3. (c) A typical Raman spectrum of HF-CVD graphene on Cu foil. Raman map of (d) 2D band intensity and (e) intensity ratio of 2D and G band from the area enclosed in the red square in (b). [Figures adapted from Ref. 27]

In this method, a tungsten or Mo filament is heated to temperatures as high as 2000 °C by passing current through it. A mixture of $CH_4$ and $H_2$ gas is showered on the hot filament through a gas shower and Cu foil is kept just below the filament. The hot temperature zone near the filament decomposes the $CH_4$ gas and heats the substrate to high temperatures of about 1000 °C. The substrate temperature depends on the position of substrate from the hot filament, i.e. the farther it is, the lower the temperature is. The advantage of this technique is that cooling rates can be very high and the Cu foil can be brought to room temperature in few minutes as it is being radiatively heated by the hot filament. This is already an industrial technique to grow large area diamond and carbon nanotube films, and it can therefore potentially be scaled up for graphene growth as well.

Thermal CVD techniques use high temperatures to decompose $CH_4$ gas into reactive radicals. However, the thermal budget (temperatures and process times) is incompatible with CMOS technology. Hence, there is a need to grow graphene at low temperatures and ideally directly on dielectric substrates, which would make the process fully compatible with the present CMOS technology. To achieve this, plasma enhanced CVD (PECVD) techniques have been proposed. Plasma assists in dissociating $CH_4$ gas into reactive radicals at low temperatures and the substrate temperatures can be kept as low as 600 °C. Fig. 3 shows a typical PECVD set up where a radio-frequency (RF) or microwave source is used to generate the plasma. The main steps of graphene growth are similar to thermal CVD with the only difference that plasma is used instead of temperature to dissociate hydrocarbon into activated carbon radicals. As with thermal CVD, the carbon radicals reach the Cu surface and form graphene layers.

Microwave and RF plasma have been used to synthesize graphene at low temperatures [29,30]. Researchers have also used remotely ignited RF plasma to grow large-area graphene on catalyst coated dielectric substrates [31]. This was done in order to avoid damage to the growing graphene layer by ion bombardment. As a downside, this process requires high vacuum conditions to operate. Although graphene films can be grown at moderate temperatures using PECVD technique, the quality of the graphene is not as good as that of thermal CVD graphene: the low process temperatures result in reduced graphene grain sizes due to restricted surface diffusion of carbon atoms on the catalytic surface. Nevertheless, due to its enhanced process compatibility, PECVD grown graphene has great promise for electronic and optoelectronic applications where highly conducting and transparent layers are required.

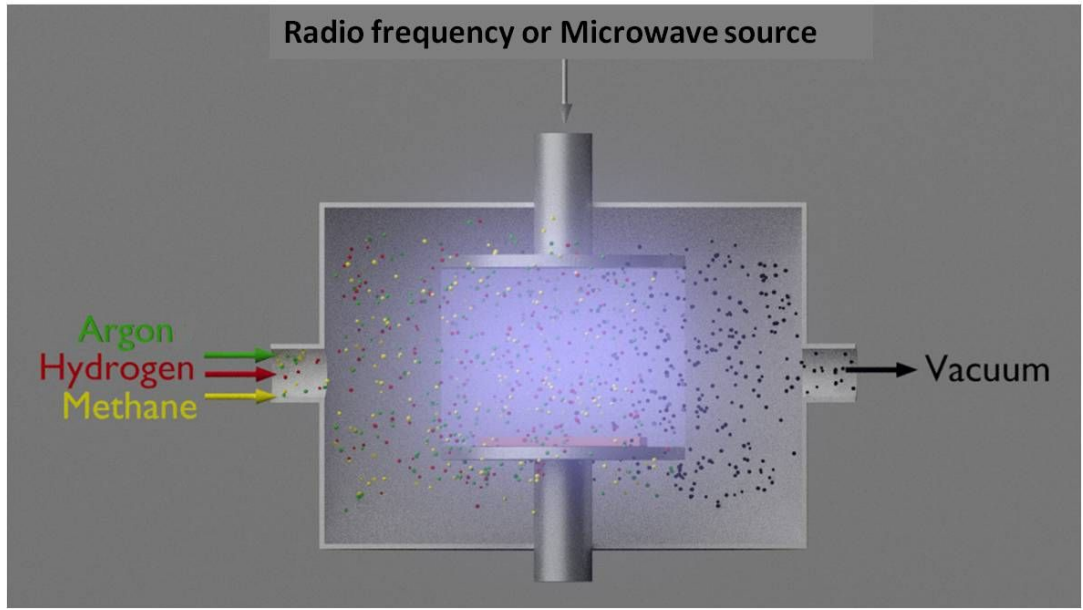

Figure 3. PECVD set up for graphene growth.

## 3. Graphene transfer

As discussed above, there have been serious efforts in growing high quality graphene by different methods. However, an often underestimated yet crucial step for almost any envisioned application of graphene lies in its efficient large-area transfer from catalytic substrates to the desired substrates, which vary according to the applications. For example, for transistor applications, graphene should be placed on a dielectric substrate. For touch screen displays and flexible devices, graphene needs to be transferred to flexible substrates like PTFE etc. The most commonly employed methods for graphene transfer are discussed below with their pros and cons.

The method proposed first for graphene transfer involves etching of the metal substrate and is a widely used wet transfer method [18]. Fig. 4(a) shows the schematic of the steps involved. First, a polymer is spin coated onto the as grown graphene on Cu and is then hot-baked to evaporate the solvent. Cu foil is then placed in an etchant solution to etch away the Cu. Most common Cu etchants are iron chloride, sodium persulphate, ammonium persulphate and HCl solution. After the Cu is etched, the polymer/graphene stack is fished out using a glass substrate or silicon wafer and placed in a water bath several times until the etchant is removed. The polymer/graphene stack is then transferred onto the desired substrate before the polymer is removed using an appropriate solvent. Fig. 4(b) shows an optical micrograph of graphene transferred onto a $SiO_2$ substrate using the wet etching transfer method. Different recipes are being used to remove polymers. PMMA is the most commonly used polymer and is generally removed using acetone. However, since it has been demonstrated that PMMA is typically not removed entirely even after long exposures to acetone, a thermal annealing step is often used to completely remove PMMA, with temperatures between 300°C and 500°C in the presence of forming gas. Considering the difficult in completely removing the PMMA residues, poly carbonate is used as an alternative that leaves less residue than PMMA [32].

A thermal release tape method is often used to transfer graphene on flexible substrates [33]. In this method, a thermal release tape is laminated on the graphene/Cu stack. After laminating, Cu is etched using a wet etching method as described above and the remaining tape/graphene stack is transferred to a flexible substrate like PTFE or PET. A pressure is applied to the tape/graphene/PET stack and then the thermal tape is released under heat. This method has been adopted for demonstrating roll-to-roll production of 30 inch graphene films for transparent electrodes [33]. The main disadvantage of this method is the problem of unwanted tape residues, which can degrade the quality of graphene. To avoid such problems, the method has been modified for direct transfer of CVD graphene onto flexible substrates [34]. In this method, the target flexible substrate is directly pressed against graphene/Cu/graphene stack and a protective sheet of weighing paper is put on top of this stack. Then the stack is sandwiched between two polyethylene terephthalate films instead of using thermal release tape.

Another recent graphene transfer method is the "soak and peel" method [35]. This method does not use corrosive etchants and hence reduces the unintentional doping of graphene by etchant ions, as well as reduces the environmental concerns. A polymer support layer is first spin coated on a graphene/Cu stack. For handling purposes, Kapton tape is then applied with pressure on the polymer support. The Cu foil is then immersed in hot de-ionized water at 90 °C for 2 hours, as shown in Fig. 4(c). In this process, water slowly seeps through the weak graphene-Cu interface and thereby slowly removes the Kapton tape / polymer / graphene stack. The Kapton tape is then pressed gently onto dielectric substrates like $SiO_2$ and is heated to around 140 °C for about 40 min, after which it can be carefully removed. Then the polymer is removed using acetone. However, this process also leaves residues on the substrate, requiring annealing steps as described above. This process, though it avoids the use of etchant solution, produces more cracks, ripples and folding and results in a non-uniform transfer, as shown by the optical micrograph in Fig. 4(d).

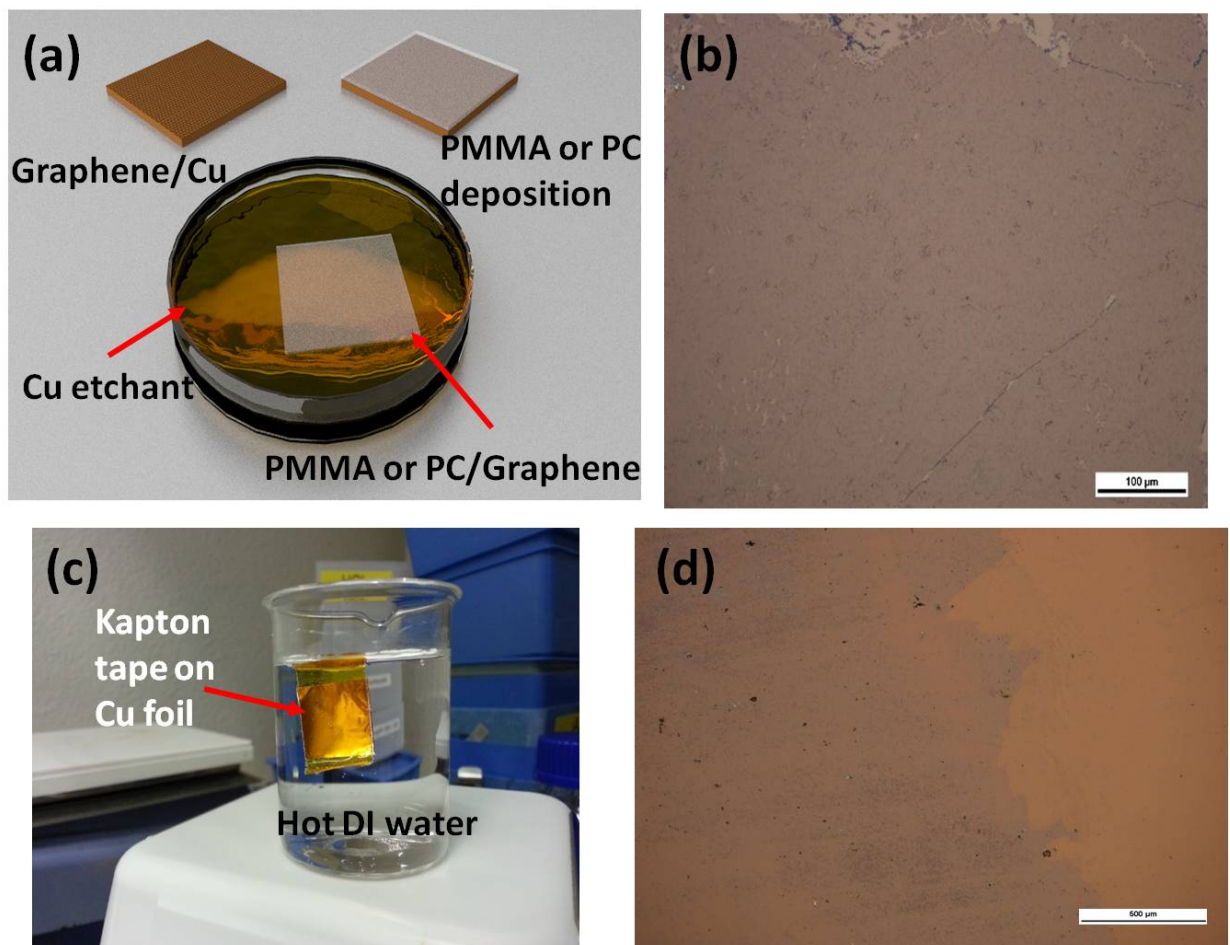

Figure 4. (a) Schematic representation of wet etching process steps. (b) Optical micrograph of graphene transferred onto a SiO$_2$ substrate using a wet etching process. (c) Photograph of a "soak and peel" process with Kapton tape. (d) Optical micrograph of Soak and Peel transferred graphene.

All the above mentioned transfer methods take from 30 minutes to few hours in removing graphene from the Cu surface. Electrochemical delamination or more commonly known as 'bubble transfer method' is a very fast and efficient method to remove graphene from a metal catalyst surface [36,37]. Like other methods, a polymer supporting layer is spin coated on the Cu surface after graphene growth. In the delamination process, a direct current (DC) voltage is applied to the polymer/Cu electrode and another electrode like Platinum and glassy carbon in an electrolyte cell arrangement, as shown in Fig. 5(a). The electrolyte solutions includes Na or K salts like NaOH, KOH, NaCl and KCl etc. A negative voltage is then applied to the Cu foil, which causes the generation of H$_2$ bubbles at the graphene-Cu interface due to the water-splitting process:

$$2H_2O(l) \longrightarrow H_2(g) + 2OH^-(aq.)$$

$H_2$ bubbles assist in removing graphene from the Cu surface, and hence the process is called bubble transfer process. Fig. 5(b) and 5(c) shows optical and scanning electron micrographs of a bubble transferred CVD graphene on $SiO_2$ substrate. Raman analysis shows that the quality of the graphene is maintained by this transfer method (Fig. 5(d)). The presence of graphene adlayers and grain boundaries, which act as defects are formed during graphene growth, resulted in a small D peak in the Raman spectrum. This process takes only a few seconds to completely remove the graphene layer from Cu surface. It is further non-destructive in nature as the metal catalyst can be reused for graphene growth many times. The method thus has potential for industrial scaling for large-area graphene transfer.

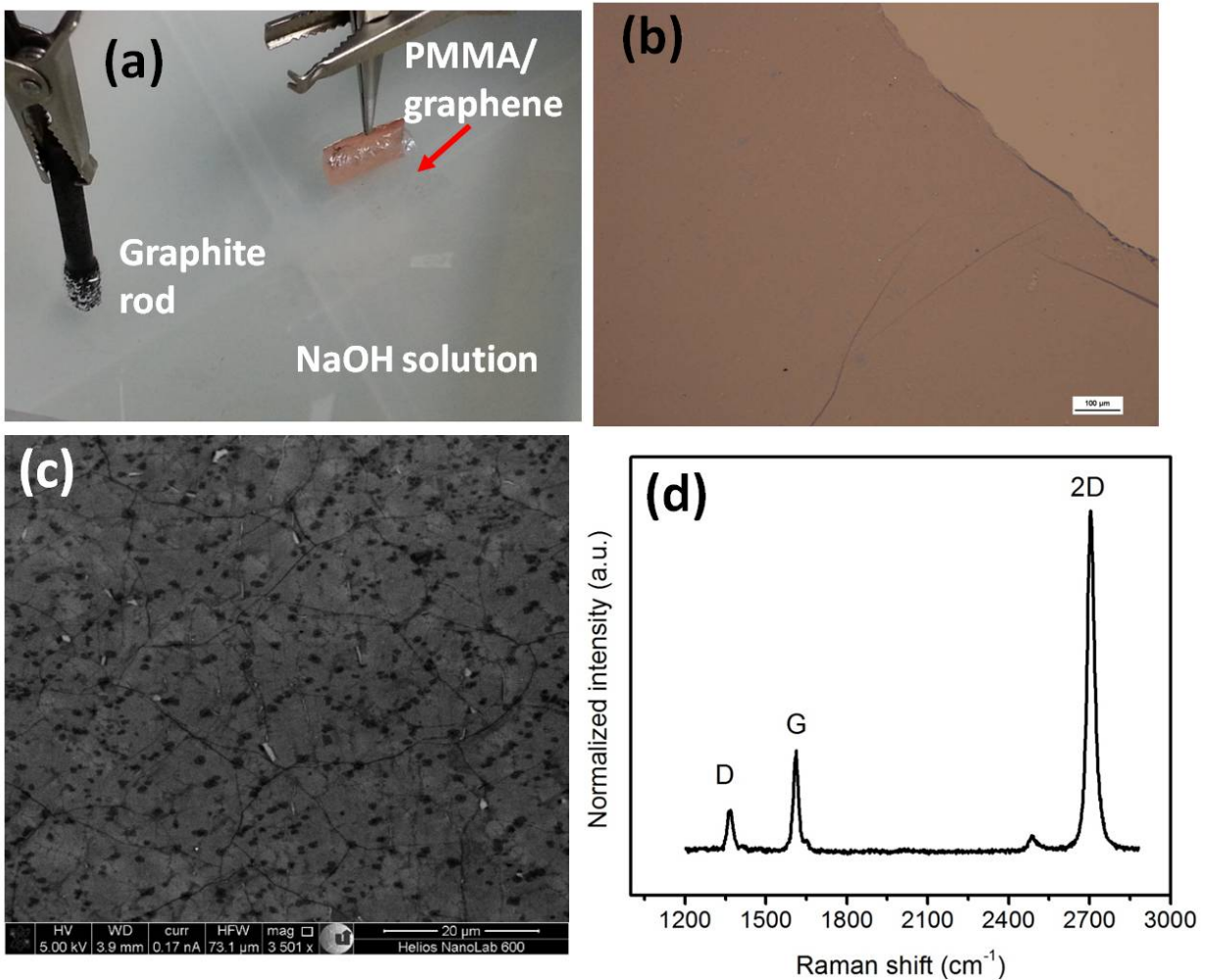

Figure 5. (a) Photograph of a typical graphene bubble transfer set up. (b) Optical micrograph of transferred graphene on $SiO_2$ substrate. (c) Scanning electron micrograph of transferred graphene. Dark spots are the graphene adlayers. (d) Raman spectra of bubble transferred graphene.

## 4. Applications

Technologically, the CVD method with its metallic catalysts and transfer processes are not (yet) compatible with semiconductor process lines. This is particularly true for the front-end-of-the-line, where stringent requirements exist regarding organic, metallic and particle contamination levels. Nevertheless, many potential applications of graphene are envisioned in microelectronics [8], optoelectronics [9] and nanoelectromechanical systems (NEMS) [10]. These applications potentially make use of the exceptional material properties of graphene, and could be integrated in the back-end-of-the-line on top of existing CMOS circuits to add "more than Moore" functionalities.

One of the first suggested applications of graphene was as a replacement channel material in field effect transistors (GFETs) to exploit the material's high carrier mobility [38]. However, the absence of an electronic band gap results in ambipolar switching characteristics and limits the current modulation in GFETs. Gapless graphene is therefore unsuitable for logic circuits, where transistors are required to display a high ratio between currents in the ON-state versus the OFF-state, i.e. they need to be switched off. Transistors for radio frequency (RF) applications, in contrast, do not require such high ON/OFF ratios [39]. As a consequence, cut-off frequencies $f_T$ (where the unilateral power gain equals unity) of up to several hundred GHz have been reported (after de-embedding) [40], rivaling those of state-of-the-art semiconductor devices. The message becomes mixed, however, once the other figure of merit for RF transistors is considered i.e. the maximum frequency of oscillation $f_{max}$. Similar to the case of logic operation, Physics strikes again. Here, the ambipolar nature of graphene leads to limited saturation in the output characteristics ($I_d$ vs. $V_d$) of GFETs, which in turn limits $f_{max}$. Therefore, bandgap engineering in graphene may be an option, either through confining the channel in one dimension with graphene nanoribbons (GNRs) [41], using electrically biased bilayer graphene [42] or by doping graphene with boron and nitrogen [43]. All of these methods would improve some of the discussed issues, but at the same time introduce new challenges to devices engineers such as random GNR orientations, GNR width fluctuations or increased design complexity due to double-gate operation or decrease in mobility after doping. Nevertheless, GFET applications remain a topic of intense research. Electronic device options beyond conventional field effect transistors include the barristor based on a graphene – silicon Schottky barrier [44], the theoretically proposed bilayer pseudospin FET (BISFET) based on an electron-hole-pair condensate [45] and lateral [46] and vertical tunnel transistors [47,48]. One such implementation, a vertical hot electron transistor, is described in more detail in sec. 4.1.

A variety of optoelectronic devices based on graphene have been demonstrated to date, building on the broadband universal absorption of 2.3% of incoming light and the high carrier mobility. While the former provides easy access to the commercially highly attractive infrared (IR) range, the latter adds potential for very high operating speeds. Graphene photodetectors typically rely on a graphene pn-junction in a diode [49] or transistor configuration [50]. They are generally extremely broadband due to the absence of a bandgap, covering a range from ultraviolet (UV) to far IR and THz, but suffer from limited responsivity due to the thinness of the material. There are several options to increase the responsivity, such as the co-integration of graphene with metallic plasmonic nanostructures [51], photon absorbing nanoparticles [52] or microcavities with built-in reflectors for specific wavelengths . All approaches are capable of enhancing the photoresponse considerably, but only for specific wavelengths. When integrated into photonic waveguides, graphene photodetectors can be quite efficient [53,54], because the light is detected in-plane with the graphene, and the graphene detector can be scaled up in size. Obviously, the RC delay of such a detector may limit the ultimate detector speed. Graphene can further act as a modulator for photonic waveguides. When gated, the Fermi level in a graphene modulator can be tuned to suppress absorption through Pauli blocking [55]. This effectively quenches absorption and results in the possibility of modulating the light passing through a (silicon) waveguide.

Micro- and nanoelectromechanical systems often employ thin membranes of various materials to form mass or pressure sensors or accelerometers and actuators. Graphene is the ultimate membrane with its ultimate thinness, record high Young's modulus of over 1 TPa [5,56], and stretchability of over 20% [57]. As a consequence, graphene based resonators can operate at very high frequencies [58]. Extrapolated values for nm-scale graphene nanomembranes are of the order of 400 GHz [59], with q-factors sufficient for the detection of single hydrogen atoms. The field of graphene NEMS is probably the least developed compared with electronics and optoelectronics, but nevertheless has high potential for applications. A discussion of an integrated graphene pressure sensor based on the piezoresistive effect is described in detail in sec. 4.2.

### 4.1. Vertical Graphene Base Hot Electron Transistors

Electronic devices with vertical carrier transport including 2D-materials have received considerable attention lately. One such candidate, a graphene-based hot electron transistor has been proposed conceptually [48] and later demonstrated in experiments [60]. In this device, a graphene sheet is sandwiched between two insulators, with metals or doped semiconductors on both sides (Fig. 6(a)). Carrier transport is vertical and happens by way of quantum mechanical tunneling. In such a device, the base contact is made up of graphene (hence, Graphene Base Transistor, GBT), because the combination of high electrical conductivity and extreme thinness of the material lead to high transmission of charge carriers. Consequently, the graphene transition time is expected to be much lower than for metals, which need to be at least 10 times thicker in order to provide the same functionality. When a voltage is applied to the graphene base, the current can be modulated by several order of magnitude (Fig. 6(b)). This is because the graphene base potential modulates the tunneling barrier between the emitter and the base. Above a certain threshold, charge carriers may tunnel via the Fowler Nordheim mechanism and reach the collector by ballistic transport (Fig. 6(c)).

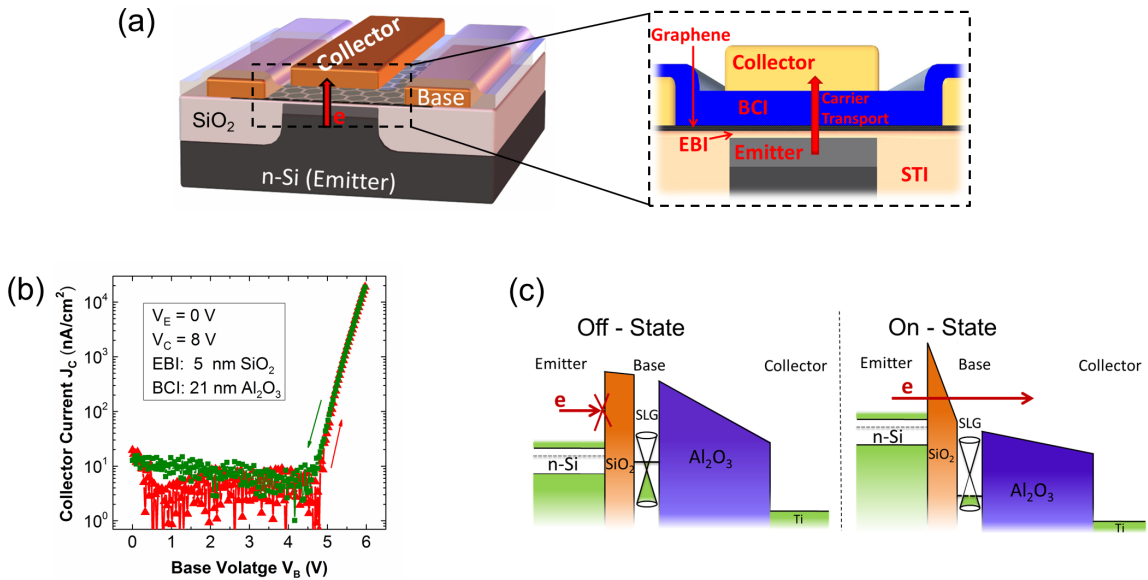

Figure 6: (a) Schematic of a vertical hot electron transistor with a base made of graphene (Graphene Base Transistor - GBT). (b) Transfer characteristics of a GBT. (c) Electronic band structure of a GBT in the off- and the on-state.

### 4.2. Graphene NEMS pressure sensor

Pressure sensors based on the piezoresistive effect in graphene membranes have been demonstrated for uniaxial strain [61] and for biaxial strain [62]. In these devices, a layer of graphene is suspended over an air-filled cavity like a drum. The device is then placed inside a vacuum chamber and air is pumped out of the chamber. As the pressure of the chamber is reduced, the pressure of the trapped air presses against the suspended graphene membrane and strains it (Fig. 7(a)). This strain changes the electronic properties of the graphene, which is measured as a change in the resistance of the graphene membrane. The device structure is shown in Fig. 7(b). Figure 7(c) shows the voltage output of one of the pressure sensors compared with the voltage output of a control device which does not have an air-filled cavity. SEM images illustrate the difference between the pressure sensor and control device. Fig. 7(d) shows the area-normalized sensitivity of a graphene pressure sensor in comparison to silicon and carbon nanotube (CNT) sensors. When normalized by unit area, the graphene sensor has a sensitivity, which is 100s of times higher than the conventional sensors. This potentially allows for aggressive device scaling and in connection with CVD graphene growth and transfer allows future on chip integration with essential cost advantages over existing technologies.

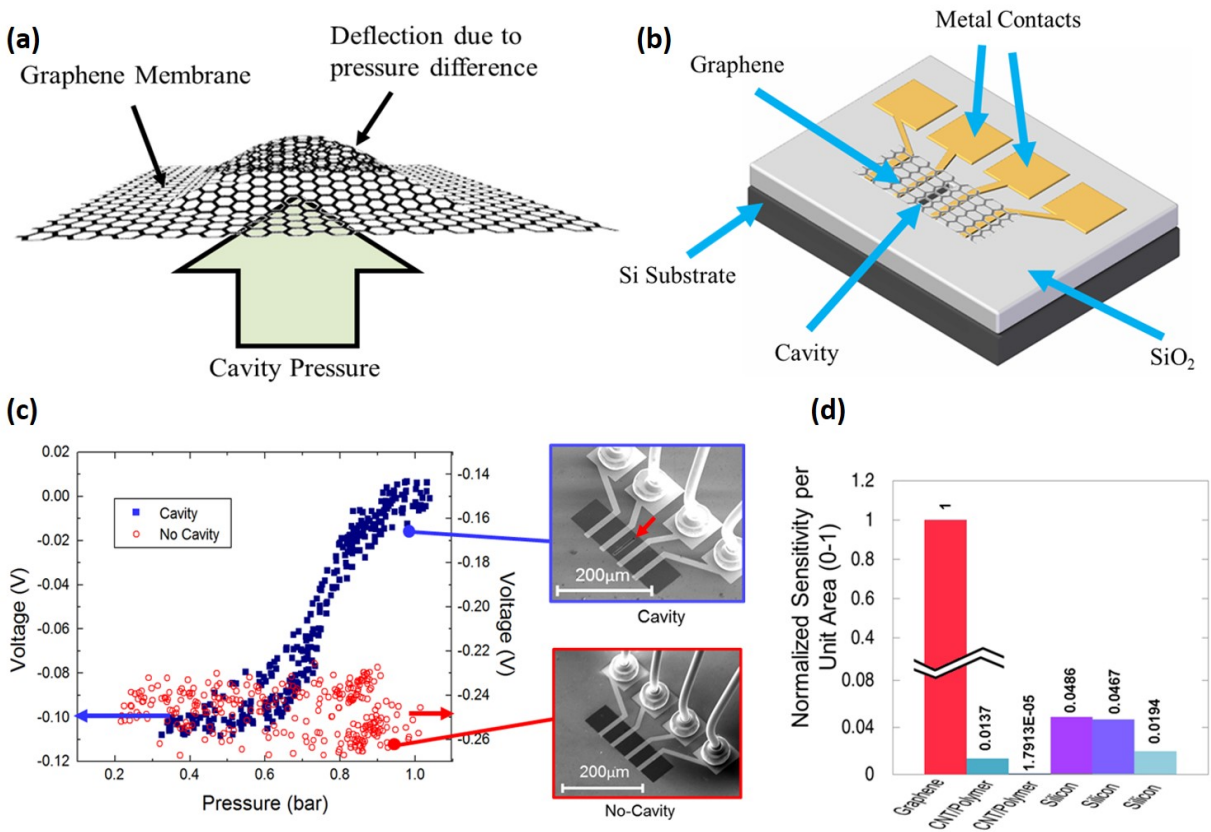

Figure 7: (a) Deflection of a graphene membrane as a result of a pressure differential between cavity pressure and the pressure of the vacuum chamber. (b) Schematic of a pressure sensor. (c) Comparison of the voltage signal output of a graphene sensor compared with a graphene device which is not suspended over an air filled cavity. SEM images illustrating the difference between devices containing a cavity (indicated by red arrow) and devices without cavities. (d) Comparison of the area-normalized sensitivities of a graphene sensor and conventional silicon and carbon nanotube sensors.

## 5. Conclusions

Graphene growth on a large-scale by CVD methods has been a key achievement on the road towards its commercialization. We discussed various CVD methods for growing graphene on different metallic substrates, especially on copper as well as very recent reports of graphene growth on germanium. Thermal CVD methods provide reasonable quality graphene but are difficult to integrate with nanoscale semiconductor processes, as high thermal budgets are needed for graphene growth. PECVD methods use low temperatures as compared to thermal CVD, but yield poor quality graphene layers. For application purposes, CVD grown graphene needs transfer to desired substrates. A number of transfer methods have, therefore, been discussed. While applications can be demonstrated with current graphene CVD technology, we conclude that further developments in graphene growth at low temperatures and direct growth on insulating substrates are still highly desired.

## Acknowledgements

Support from the European Commission through a STREP project (GRADE, No. 317839), an ERC Starting Grant (InteGraDe, No. 307311), an ERC Advanced Investigator Grant (OSIRIS, No. 228229) as well as the German Research Foundation (DFG, LE 2440/1-1) is gratefully acknowledged.

# References

[1]  K. S. Novoselov, A. K. Geim, S. V. Morozov, D. Jiang, Y. Zhang, S. V. Dubonos, I. V. Grigorieva, and A. A. Firsov, Science **306**, 666 (2004).
[2]  K. I. Bolotin, K. J. Sikes, Z. Jiang, M. Klima, G. Fudenberg, J. Hone, P. Kim, and H. L. Stormer, Solid State Commun. **146**, 351 (2008).
[3]  R. R. Nair, P. Blake, A. N. Grigorenko, K. S. Novoselov, T. J. Booth, T. Stauber, N. M. R. Peres, and A. K. Geim, Science **320**, 1308 (2008).
[4]  R. Murali, Y. Yang, K. Brenner, T. Beck, and J. D. Meindl, Appl. Phys. Lett. **94**, 243114 (2009).
[5]  C. Lee, X. Wei, J. W. Kysar, and J. Hone, Science **321**, 385 (2008).
[6]  A. A. Balandin, S. Ghosh, W. Bao, I. Calizo, D. Teweldebrhan, F. Miao, and C. N. Lau, Nano Lett. **8**, 902 (2008).
[7]  K. S. Novoselov, V. I. Fal′ko, L. Colombo, P. R. Gellert, M. G. Schwab, and K. Kim, Nature **490**, 192 (2012).
[8]  Y. Wu, D. B. Farmer, F. Xia, and P. Avouris, Proc. IEEE **101**, 1620 (2013).
[9]  F. Bonaccorso, Z. Sun, T. Hasan, and A. C. Ferrari, Nat. Photonics **4**, 611 (2010).
[10] C. Chen and J. Hone, Proc. IEEE **101**, 1766 (2013).
[11] S. Vaziri, G. Lupina, A. Paussa, A. D. Smith, C. Henkel, G. Lippert, J. Dabrowski, W. Mehr, M. Östling, and M. C. Lemme, Solid-State Electron. **84**, 185 (2013).
[12] A. D. Smith, S. Vaziri, S. Rodriguez, M. Ostling, and M. C. Lemme, in *2014 15th Int. Conf. Ultim. Integr. Silicon ULIS* (2014), pp. 29–32.
[13] C. Berger, Z. Song, T. Li, X. Li, A. Y. Ogbazghi, R. Feng, Z. Dai, A. N. Marchenkov, E. H. Conrad, P. N. First, and W. A. de Heer, J. Phys. Chem. B **108**, 19912 (2004).
[14] Q. Yu, J. Lian, S. Siriponglert, H. Li, Y. P. Chen, and S.-S. Pei, Appl. Phys. Lett. **93**, 113103 (2008).
[15] X. Li, W. Cai, J. An, S. Kim, J. Nah, D. Yang, R. Piner, A. Velamakanni, I. Jung, E. Tutuc, S. K. Banerjee, L. Colombo, and R. S. Ruoff, Science **324**, 1312 (2009).
[16] Y. Hu, M. Ruan, Z. Guo, R. Dong, J. Palmer, J. Hankinson, C. Berger, and W. A. de Heer, J. Phys. Appl. Phys. **45**, 154010 (2012).
[17] H. Zhou, W. J. Yu, L. Liu, R. Cheng, Y. Chen, X. Huang, Y. Liu, Y. Wang, Y. Huang, and X. Duan, Nat. Commun. **4**, 2096 (2013).
[18] C. Mattevi, H. Kim, and M. Chhowalla, J. Mater. Chem. **21**, 3324 (2011).
[19] Y. Hao, M. S. Bharathi, L. Wang, Y. Liu, H. Chen, S. Nie, X. Wang, H. Chou, C. Tan, B. Fallahazad, H. Ramanarayan, C. W. Magnuson, E. Tutuc, B. I. Yakobson, K. F. McCarty, Y.-W. Zhang, P. Kim, J. Hone, L. Colombo, and R. S. Ruoff, Science **342**, 720 (2013).
[20] I. Vlassiouk, M. Regmi, P. Fulvio, S. Dai, P. Datskos, G. Eres, and S. Smirnov, ACS Nano **5**, 6069 (2011).
[21] Z. Sun, A.-R. O. Raji, Y. Zhu, C. Xiang, Z. Yan, C. Kittrell, E. L. G. Samuel, and J. M. Tour, ACS Nano **6**, 9790 (2012).
[22] H. Kim, I. Song, C. Park, M. Son, M. Hong, Y. Kim, J. S. Kim, H.-J. Shin, J. Baik, and H. C. Choi, ACS Nano **7**, 6575 (2013).
[23] T. Kobayashi, M. Bando, N. Kimura, K. Shimizu, K. Kadono, N. Umezu, K. Miyahara, S. Hayazaki, S. Nagai, Y. Mizuguchi, Y. Murakami, and D. Hobara, Appl. Phys. Lett. **102**, 023112 (2013).
[24] G. Wang, M. Zhang, Y. Zhu, G. Ding, D. Jiang, Q. Guo, S. Liu, X. Xie, P. K. Chu, Z. Di, and X. Wang, Sci. Rep. **3**, 2465 (2013).
[25] G. Lippert, J. Dąbrowski, T. Schroeder, M. A. Schubert, Y. Yamamoto, F. Herziger, J. Maultzsch, J. Baringhaus, C. Tegenkamp, M. C. Asensio, J. Avila, and G. Lupina, Carbon **75**, 104 (2014).
[26] J.-H. Lee, E. K. Lee, W.-J. Joo, Y. Jang, B.-S. Kim, J. Y. Lim, S.-H. Choi, S. J. Ahn, J. R. Ahn, M.-H. Park, C.-W. Yang, B. L. Choi, S.-W. Hwang, and D. Whang, Science **344**, 286 (2014).
[27] S. Kataria, A. Patsha, S. Dhara, A. K. Tyagi, and H. C. Barshilia, J. Raman Spectrosc. **43**, 1864 (2012).
[28] R. Hawaldar, P. Merino, M. R. Correia, I. Bdikin, J. Grácio, J. Méndez, J. A. Martín-Gago, and M. K. Singh, Sci. Rep. **2**, 682 (2012).
[29] T. Yamada, J. Kim, M. Ishihara, and M. Hasegawa, J. Phys. Appl. Phys. **46**, 063001 (2013).
[30] T. Terasawa and K. Saiki, Carbon **50**, 869 (2012).
[31] Y. Wang, X. Xu, J. Lu, M. Lin, Q. Bao, B. Özyilmaz, and K. P. Loh, ACS Nano **4**, 6146 (2010).
[32] Y.-C. Lin, C. Jin, J.-C. Lee, S.-F. Jen, K. Suenaga, and P.-W. Chiu, ACS Nano **5**, 2362 (2011).
[33] S. Bae, H. Kim, Y. Lee, X. Xu, J.-S. Park, Y. Zheng, J. Balakrishnan, T. Lei, H. Ri Kim, Y. I. Song, Y.-J. Kim, K. S. Kim, B. Özyilmaz, J.-H. Ahn, B. H. Hong, and S. Iijima, Nat. Nanotechnol. **5**, 574 (2010).
[34] L. G. P. Martins, Y. Song, T. Zeng, M. S. Dresselhaus, J. Kong, and P. T. Araujo, Proc. Natl. Acad. Sci. 201306508 (2013).
[35] P. Gupta, P. D. Dongare, S. Grover, S. Dubey, H. Mamgain, A. Bhattacharya, and M. M. Deshmukh, Sci. Rep. **4**, 3882 (2014).
[36] Y. Wang, Y. Zheng, X. Xu, E. Dubuisson, Q. Bao, J. Lu, and K. P. Loh, ACS Nano **5**, 9927 (2011).
[37] T. Ciuk, I. Pasternak, A. Krajewska, J. Sobieski, P. Caban, J. Szmidt, and W. Strupinski, J. Phys. Chem. C **117**, 20833 (2013).
[38] M. C. Lemme, T. J. Echtermeyer, M. Baus, and H. Kurz, IEEE Electron Device Lett. **28**, 282 (2007).
[39] F. Schwierz, Proc. IEEE **101**, 1567 (2013).

[40] R. Cheng, J. Bai, L. Liao, H. Zhou, Y. Chen, L. Liu, Y.-C. Lin, S. Jiang, Y. Huang, and X. Duan, Proc. Natl. Acad. Sci. **109**, 11588 (2012).
[41] X. Li, X. Wang, L. Zhang, S. Lee, and H. Dai, Science **319**, 1229 (2008).
[42] B. N. Szafranek, D. Schall, M. Otto, D. Neumaier, and H. Kurz, Nano Lett. **11**, 2640 (2011).
[43] C.-K. Chang, S. Kataria, C.-C. Kuo, A. Ganguly, B.-Y. Wang, J.-Y. Hwang, K.-J. Huang, W.-H. Yang, S.-B. Wang, C.-H. Chuang, M. Chen, C.-I. Huang, W.-F. Pong, K.-J. Song, S.-J. Chang, J.-H. Guo, Y. Tai, M. Tsujimoto, S. Isoda, C.-W. Chen, L.-C. Chen, and K.-H. Chen, ACS Nano **7**, 1333 (2013).
[44] H. Yang, J. Heo, S. Park, H. J. Song, D. H. Seo, K.-E. Byun, P. Kim, I. Yoo, H.-J. Chung, and K. Kim, Science **336**, 1140 (2012).
[45] S. K. Banerjee, L. F. Register, E. Tutuc, D. Reddy, and A. H. MacDonald, IEEE Electron Device Lett. **30**, 158 (2009).
[46] J.-S. Moon, D. Curtis, S. Bui, M. Hu, D. K. Gaskill, J. L. Tedesco, P. Asbeck, G. G. Jernigan, B. L. VanMil, R. L. Myers-Ward, J. Eddy, C.R., P. M. Campbell, and X. Weng, IEEE Electron Device Lett. **31**, 260 (2010).
[47] L. Britnell, R. V. Gorbachev, R. Jalil, B. D. Belle, F. Schedin, A. Mishchenko, T. Georgiou, M. I. Katsnelson, L. Eaves, S. V. Morozov, N. M. R. Peres, J. Leist, A. K. Geim, K. S. Novoselov, and L. A. Ponomarenko, Science **335**, 947 (2012).
[48] W. Mehr, J. Dabrowski, J. Christoph Scheytt, G. Lippert, Y.-H. Xie, M. C. Lemme, M. Ostling, and G. Lupina, IEEE Electron Device Lett. **33**, 691 (2012).
[49] T. Mueller, F. Xia, and P. Avouris, Nat Photon **4**, 297 (2010).
[50] M. C. Lemme, F. H. L. Koppens, A. L. Falk, M. S. Rudner, H. Park, L. S. Levitov, and C. M. Marcus, Nano Lett. **11**, 4134 (2011).
[51] T. J. Echtermeyer, L. Britnell, P. K. Jasnos, A. Lombardo, R. V. Gorbachev, A. N. Grigorenko, A. K. Geim, A. C. Ferrari, and K. S. Novoselov, Nat Commun **2**, 458 (2011).
[52] G. Konstantatos, M. Badioli, L. Gaudreau, J. Osmond, M. Bernechea, F. P. G. de Arquer, F. Gatti, and F. H. L. Koppens, Nat. Nanotechnol. **7**, 363 (2012).
[53] A. Pospischil, M. Humer, M. M. Furchi, D. Bachmann, R. Guider, T. Fromherz, and T. Mueller, Nat. Photonics **7**, 892 (2013).
[54] X. Gan, R.-J. Shiue, Y. Gao, I. Meric, T. F. Heinz, K. Shepard, J. Hone, S. Assefa, and D. Englund, Nat. Photonics **7**, 883 (2013).
[55] M. Liu, X. Yin, E. Ulin-Avila, B. Geng, T. Zentgraf, L. Ju, F. Wang, and X. Zhang, Nature **474**, 64 (2011).
[56] G.-H. Lee, R. C. Cooper, S. J. An, S. Lee, A. van der Zande, N. Petrone, A. G. Hammerberg, C. Lee, B. Crawford, W. Oliver, J. W. Kysar, and J. Hone, Science **340**, 1073 (2013).
[57] H. Tomori, A. Kanda, H. Goto, Y. Ootuka, K. Tsukagoshi, S. Moriyama, E. Watanabe, and D. Tsuya, Appl Phys Express **4**, 075102 (2011).
[58] A. Eichler, J. Moser, J. Chaste, M. Zdrojek, I. Wilson-Rae, and A. Bachtold, Nat. Nanotechnol. **6**, 339 (2011).
[59] T. Mashoff, M. Pratzer, V. Geringer, T. J. Echtermeyer, M. C. Lemme, M. Liebmann, and M. Morgenstern, Nano Lett. **10**, 461 (2010).
[60] S. Vaziri, G. Lupina, C. Henkel, A. D. Smith, M. Östling, J. Dabrowski, G. Lippert, W. Mehr, and M. C. Lemme, Nano Lett. **13**, 1435 (2013).
[61] A. D. Smith, F. Niklaus, A. Paussa, S. Vaziri, A. C. Fischer, M. Sterner, F. Forsberg, A. Delin, D. Esseni, P. Palestri, M. Östling, and M. C. Lemme, Nano Lett. **13**, 3237 (2013).
[62] A. D. Smith, F. Niklaus, S. Vaziri, A. C. Fischer, M. Sterner, F. Forsberg, S. Schroder, M. Ostling, and M. C. Lemme, in *2014 IEEE 27th Int. Conf. Micro Electro Mech. Syst. MEMS* (2014), pp. 1055–1058.